\documentclass[11pt]{article}
\usepackage{amsmath}
\begin{document}

\thispagestyle{empty}

\begin{flushright} LPTENS 0419 \end{flushright}

\vspace*{0.5cm}

\begin{center}{\LARGE { GAUGE THEORIES FORMULATED ON 
\vskip 0.4cm
SURFACES WITH NON-COMMUTATIVE 
\vskip 0.5cm
GEOMETRY}}

\end{center}

\vskip1cm
\begin{center}
{\bf{E.G. Floratos}}

\vskip0.2cm
Physics Dept. University of Athens, Greece\\
floratos@inp.demokritos.gr

\vskip0.5cm
{\bf{J. Iliopoulos}}
\vskip0.2cm

Laboratoire de Physique Th\'eorique CNRS-ENS\\ 
24 rue Lhomond, F-75231 Paris Cedex 05, France\\
ilio@lpt.ens.fr
\vskip1.0cm

{\bf ABSTRACT}

\end{center}

\bigskip

We propose a formulation of $d$-dimensional classical $SU(N)$ Yang-Mills theories on a
$d+2$-dimensional space, with the extra two dimensions forming a surface with
non-commutative geometry. This equivalence is valid in any finite order in the $1/N$ expansion.

\bigskip

\newpage

\section{Introduction}

For theories which do not possess a natural small expansion parameter, the inverse
of the number of variables has often been used at the limit when this number becomes very large. 
The simplest example is provided by an $N$-component real scalar field with an $O(N)$ invariant 
interaction:

\begin{equation}
\label{lint}
 {\cal L}_{int} = -\lambda \biggl( \sum_{i=1}^N \phi^{i} \phi^{i} \biggr)^2 
\end{equation}

As $N$ grows the diagrams with the largest power of $N$ dominate. The result \cite{wilson}, 
\cite{coleman},  is that the model is soluble at the limit :

\begin{equation}
\label{ONlim}
N \rightarrow \infty ~~~~~~~~{\rm keeping}~~~ \lambda N~{\rm fixed}
\end{equation}
Notice that, in this limit $\lambda$ goes to zero.

The simplicity of this result is due to the fact that, when $N$ goes to infinity, only a very small 
subset of all the diagrams, the so-called "sausage diagrams", survive. They
can be summed explicitly. The situation changes if we turn to Yang-Mills
theories based on the group $SU(N)$ \cite{hooft}, \cite{venez}. In this case G. 't Hooft
showed that the right limit to consider is

\begin{equation}
\label{hooftlim}
N \rightarrow \infty ~~~~~~~~{\rm keeping}~~~ g^2 N~{\rm fixed}
\end{equation}
where $g$ is the gauge theory coupling constant. Contrary to the previous case the subset of dominant diagrams contains all the planar ones. Nobody has succeeded in summing them, so the model is not soluble. On the other hand, precisely because the dominant diagrams form a much richer set, one hopes that the large $N$ limit contains already all the essential non-perturbative properties of non-abelian Yang-Mills theories, such as confinement \cite{witten}. It is straightforward to extend this result to all orders in an expansion in powers of $1/N$. The first correction to the above limit is given by the sum of all diagrams which form a surface with one handle, the second with two handles ...etc.

The purpose of this note is to state and, to a certain extend, prove, the following statement:

\vskip0.5cm

{\it Statement:} Given an $SU(N)$ Yang-Mills theory  in a $d-$dimensional space
with potentials

\begin{equation}
\label{gaugepot}
 A_{\mu}(x)~=~A_{\mu}^a (x)~t_a 
\end{equation}
where $t_a$ are the standard $SU(N)$ matrices, there exists a reformulation in which 
the gauge fields and the gauge potentials become:

\begin{equation}
\label{newlimits}
A_{\mu}(x) \rightarrow {\cal A}_{\mu}(x,z_1,z_2)~~~~~~~ F_{\mu \nu}(x) \rightarrow {\cal F}_{\mu \nu}(x,z_1,z_2)
\end{equation}
where ${\cal A}$ and ${\cal F}$ are fields in a $(d+2)-$dimensional space, greek indices still run from 0 to $d-1$ and $z_1$ and $z_2$ are
local coordinates on a  two-dimensional surface endowed with non-commutative
geometry \cite{connes}. They will be shown to satisfy the commutation relation

\begin{equation}
\label{fuzsphcom}
[z_1,z_2]=\frac{2i}{N}
\end{equation}

The commutators in  the original $SU(N)$ Yang-Mills theory are
replaced by the Moyal brackets \cite{moyal}, \cite{fairlie} with respect to the non-commuting coordinates.

\begin{equation}
\begin{split}
\label{commoy}
[A_{\mu}(x), A_{\nu}(x)] & \rightarrow \{{\cal A}_{\mu}(x,z_1,z_2), {\cal
    A}_{\nu}(x,z_1,z_2)\}_{Moyal}\\
[A_{\mu}(x),\Omega(x)] & \rightarrow \{{\cal A}_{\mu}(x,z_1,z_2), 
 {\it   \Omega} (x,z_1,z_2)\}_{Moyal}
\end{split}
\end{equation}
where $\Omega $ is the function of the gauge transfomation and $\{,\}_{Moyal}$
denotes the Moyal bracket with respect to the two operators $z_1$ and $z_2$.
The trace over the group indices in the original Yang-Mills action becomes a
two dimensional integral over the surface:

\begin{equation} 
\label{action}
\int d^4x~ Tr\left( F_{\mu \nu}(x)F^{\mu \nu}(x)\right)~~\rightarrow~~ \int d^4xdz_1dz_2~{\cal F}_{\mu \nu}(x,z_1,z_2)*{\cal F}^{\mu \nu}(x,z_1,z_2)
\end{equation}

The *-product will be defined later. When $N$ goes to infinity, the two $z$'s
commute and the *-product reduces to the ordinary product.

In what follows we shall give a
partial proof of this statement.
\vskip 0.5 cm

Field theories on spaces with non-commutative geometry have been studied extensively in recent years \cite{douglas}, \cite{jp}, \cite{sw}, \cite{elias}. 

The article is organized as follows: In section 2 we set up the framework,
define an appropriate $1/N$ expansion and give an algebraic proof of the
statement in the leading $1/N$ order. In section 3 we establish the
non-commutative geometry and show how the Moyal bracket emerges to all orders
in $1/N$. Finally, section 4 contains some remarks and our
conclusions. 

\section{The large $N$ limit}

The large $N$ limit we shall consider in this section can be viewed as dual to the one
introduced by 't Hooft in eq. (\ref{hooftlim}), namely we shall take the strong coupling limit

\begin{equation}
N \rightarrow \infty ~~~~~~~~{\rm keeping}~~~ g^{-2} N^{3}~{\rm fixed}
\label{newlimit}
\end{equation}

It is the limit in which the correspondance between Yang-Mills theories and
matrix models has been established.
\vskip 0.5 cm 

Before going into the proof of the statement, let us recall some
well-known results from the theory of
membranes \cite{duff}.

The closed bosonic membrane is described by the world-volume Lagrangian:

\begin{equation}
\begin{split}
L~=&T\sqrt{g}  \\
g~=&detg_{\alpha \beta},~~~~g_{\alpha \beta}~=\partial_{\alpha }X^{\mu }\partial_{\beta }X_{\mu }
\label{Lmembr}
\end{split}
\end{equation}

Here $X_{\mu }(\tau ,\sigma_{1} ,\sigma_{2})$, $\mu =0,1,...,d-1$ is the
membrane embedding
in a $d-$dimensional ambient Minkowski space-time and $g_{\alpha \beta}$, $\alpha ,\beta =
\tau , \sigma_{1} ,\sigma_{2}$ is the induced metric on the world-volume. $T$ is the membrane 
tension.

The action corresponding to the Lagrangian (\ref{Lmembr}) is invariant under reparametrisations of the world-volume. After 
the light-cone gauge fixing, the symmetry transformations which remain are the area-preserving 
diffeomorphisms
of the surface:

\begin{equation}
\begin{split}
\sigma_{1} ~,~\sigma_{2} &~\rightarrow ~\tilde{\sigma}_{1}(\sigma_{1} ,\sigma_{2}) ~,~\tilde{\sigma}_{2}(\sigma_{1} ,\sigma_{2}) \\
det~& \frac {\partial{(\tilde{\sigma}_{1}, \tilde{\sigma}_{2})}}{\partial{(\sigma_{1} ,\sigma_{2})}} ~ =~1
\label{arpres}
\end{split}
\end{equation}

For an infinitesimal diffeomorphism $\delta \sigma_{1}=u_{1}$ and $\delta \sigma_{2}=u_{2}$, area preservation is equivalent to the condition :

\begin{equation}
\label{condition}
\frac {\partial{u_{1}}}{\partial{\sigma_{1}}}+\frac {\partial{u_{2}}}{\partial{\sigma_{2}}}=0
\end{equation}

A particular solution of (\ref{condition}) is obtained by expressing the two components of the vector field $u=(u_{1},u_{2})$ in terms of a single scalar function $h(\sigma_{1} ,\sigma_{2})$:

\begin{equation}
\label{partsol}
u_{1}=\frac {\partial{h}}{\partial{\sigma_{2}}}  ,~~~~u_{2}=-\frac {\partial{h}}{\partial{\sigma_{1}}}
\end{equation} 

For the simplest case of the sphere $S^{2}$, (\ref{partsol}) provides the most
general solution. Choosing $\sigma_{1} =\phi $ and $\sigma_{2} =cos\theta $, we can expand $h$ on the basis of the spherical harmonics  $Y_{l,m} (\theta ,\phi )$. The generators of the area-preserving diffeomorphisms can then be expressed as:

\begin{equation}
\label{sphgener}
L_{l,m}=\frac {\partial{Y_{l,m}}}{\partial{ cos\theta}} \frac {\partial{}}{\partial{\phi}}-
\frac {\partial{Y_{l,m}}}{\partial{ \phi}} \frac {\partial{}}{\partial{cos\theta}}
\end{equation}

They satisfy the Lie algebra:

\begin{equation}
\label{sphalg}
[L_{l,m},L_{l',m' }]=f^{l'',m''}_{l,m ;~ l',m'} L_{l'',m''}
\end{equation}
with the structure constants given by:

\begin{equation}
\label{sphstrcts}
\{Y_{l,m},Y_{l',m'} \}=f^{l'',m''}_{l,m ;~ l',m'}Y_{l'',m''}
\end{equation}
where the curly bracket represents the Poisson bracket with respect to $\phi $ and
$cos\theta $ \cite{hoppe}, \cite{fi1}. Similar results hold for the torus \cite{fi1}, \cite{zachos1}.

\vskip0.5cm

After these preliminaries, we turn to the proof of our statement. In order to
be specific, let us concentrate on the case of the 
sphere. The main step of the proof is algebraic, namely we shall show that the Lie algebra of
the group $SU(N)$, at the limit when $N$ goes to infinity, with the generators appropriately 
rescaled, becomes the algebra of the area preserving diffeomorphisms of the sphere 
(\ref{sphalg}). It is possible to construct a direct proof at the level of the structure constants 
of the algebra. In particular, one can show that the structure constants of $[SDiff(S^2)]$
given by (\ref {sphstrcts}) are the limits for large $N$ of those of $SU(N)$. Such proofs exist
already in the literature \cite{hoppe}. Here we shall present an alternative proof which may give a new insight 
to the problem \cite{fi2}. We first remark that the spherical harmonics $Y_{l,m}(\theta ,\phi)$ are
harmonic homogeneous polynomials of degree $l$ in three euclidean coordinates $x_{1}$, 
$x_{2}$, $x_{3}$:

\begin{equation}
\label{sphcoord}
x_{1}=cos\phi ~sin\theta,~~~~x_{2}=sin\phi ~sin\theta, ~~~~x_{3}=cos\theta
\end{equation}

\begin{equation}
\label{Ylm}
Y_{l,m} (\theta, \phi)=~~\sum _{i_{k}=1,2,3 \atop k=1,...,l}
\alpha_{i_{1}...i_{l}}^{(m)}~x_{i_{1}}...x_{i_{l}}
\end{equation}
where $\alpha_{i_{1}...i_{l}}^{(m)}$ is a symmetric and traceless tensor. For fixed $l$ there are 
$2l+1$ linearly independent tensors $\alpha_{i_{1}...i_{l}}^{(m)}$, $m=-l,...,l$. 

Let us now choose, inside $SU(N)$, an $SU(2)$ subgroup by choosing three $N\times N$ hermitian 
matrices which form an $N-$dimensional irreducible representation of the Lie algebra of $SU(2)$. They correspond to the non-commutative coordinates of a fuzzy sphere \cite{madore}

\begin{equation}
\label{su2}
[S_{i},S_{j}]=i\epsilon_{ijk}S_{k}
\end{equation}

The $S$ matrices, together with the $\alpha$ tensors introduced before, can be
used to construct a basis of $N^2-1$ matrices acting on the fundamental
representation of $SU(N)$ \cite{schwinger}. 

\begin{equation}
\begin{split}
S^{(N)}_{l,m}=~~\sum _{i_{k}=1,2,3 \atop k=1,...,l}
\alpha_{i_{1}...i_{l}}^{(m)}~S_{i_{1}}...S_{i_{l}}  \\
[S^{(N)}_{l,m},~S^{(N)}_{l',m'}]=if^{(N)l'',m''}_{l,m ;~ l',m'}~S^{(N)}_{l'',m''}
\label{sunalg}
\end{split}
\end{equation}
where the $f'$s appearing in the r.h.s. of (\ref {sunalg}) are just the $SU(N)$ structure constants in a 
somehow unusual notation. Their normalisation is given by

\begin{equation}
\label{trace}
Tr\left(S^{(N)}_{l,m}~S^{(N)}_{l',m'}\right)~=~\frac {1}{4\pi} N\left(\frac {N^2-1}{4}\right)^l~\delta_{ll'}\delta_{mm'}
\end{equation}

The important, although trivial, observation is that the three $SU(2)$ generators $S_{i}$, 
rescaled by a factor proportional to $1/N$, will have well-defined limits as
$N$ goes to infinity.

\begin{equation}
\label{rescsu2gen}
S_{i}\rightarrow T_{i} = \frac {2}{N} S_{i}
\end{equation}

Indeed, all matrix elements of $T_i$ are bounded by $|(T_i)_{ab}|\leq$ 1. 
They satisfy the rescaled algebra:

\begin{equation}
\label{rescsu2}
[T_{i},T_{j}]=\frac {2i}{N} \epsilon _{ijk}T_{k}
\end{equation}
and the Casimir element

\begin{equation}
\label{casimir}
T^2=T_{1}^2+T_{2}^2+T_{3}^2=1-\frac {1}{N^2}
\end{equation}
in other words,  under the norm $\| x \| ^2 = Trx^2 $, the limits as $N$ goes to infinity of the 
generators $T_{i}$ are three objects $x_{i}$ which commute by (\ref{rescsu2}) and are constrained 
by (\ref{casimir}):

\begin{equation}
\label{constr}
x_{1}^2+x_{2}^2+x_{3}^2=1
\end{equation}

If we consider two polynomial functions $f(x_{1},x_{2},x_{3})$ and $g(x_{1},x_{2},x_{3})$ the corresponding matrix polynomials  $f(T_{1},T_{2},T_{3})$ and $g(T_{1},T_{2},T_{3})$ have commutation relations for large $N$ which follow from (\ref{rescsu2}):

\begin{equation}
\label{limpois}
\frac {N}{2i}~ [f,g] \rightarrow ~~\epsilon_{ijk} ~x_{i}~ \frac {\partial {f}}{\partial {x_{j}}} \frac {\partial {g}}{\partial {x_{k}}} 
\end{equation}

If we replace now in the $SU(N)$ basis (\ref {sunalg}) the $SU(2)$ generators $S_{i}$ by the rescaled ones $T_{i}$, we obtain a set of $N^2-1$ matrices $T^{(N)}_{l,m}$ which, according to (\ref{Ylm}), (\ref{sunalg}) and (\ref{limpois}), satisfy:

\begin{equation}
\label{limpoisY}
\frac {N}{2i}~ [T^{(N)}_{l,m},T^{(N)}_{l',m'}] \rightarrow ~\{ Y_{l,m},Y_{l',m'} \}
\end{equation}

The relation (\ref{limpoisY}) completes the algebraic part of the proof. It shows that the 
$SU(N)$ algebra, under the rescaling (\ref{rescsu2gen}), does go to that of $[SDiff(S^2)]$. 
It is now straightforward to obtain the limit for the Yang-Mills action. 
We expand the classical fields $SU(N)$ in the basis of the matrices $T^{(N)}_{l,m}$:

\begin{equation}
\label{gaugefield}
A_{\mu}(x)=~~\sum_{l=1,...,N-1\atop m=-l,...,l} 
A_{\mu}^{l,m}(x) T^{(N)}_{l,m} 
\end{equation}
which, according to (\ref{limpoisY}), implies that the rescaled commutator has as a limit 
the Poisson bracket:

\begin{equation}
\label{finallimitsph}
N[A_{\mu},A_{\nu}] \rightarrow~~~\{ A_{\mu}(x,\theta ,\phi ), A_{\nu}(x,\theta ,\phi ) \}
\end{equation}

Combining (\ref{finallimitsph}) with the fact that the trace of the $T$ matrices is obtained from 
(\ref{trace}) by rescaling:

\begin{equation}
\label{resctrace}
Tr\left(T^{(N)}_{l,m}~T^{(N)}_{l',m'}\right)~\rightarrow N~\delta_{ll'}\delta_{mm'}
\end{equation}
we obtain the result for the case of the sphere. The torus can be treated similarly \cite{zachos2}. For a general discussion on the correspondance between gauge theories and $d+2$ dimensional local fields, see \cite{sakita}.

\section{Higher orders}

In this section we shall argue that the equivalense between Yang-Mills
theories and field theories on surfaces is in fact valid to any order in
$1/N$. The new feature is that the $SU(N)$ algebra induces a non-commutative
geometry on the surface.

\vskip 0.5 cm

In order to be specific, let us come back to the case of the sphere. As shown
in the previous section, we can choose an $SU(2)$ basis for any
finite $N$ and the rescaled algebra (\ref{rescsu2}) and (\ref{casimir}) holds
exactly, without any higher order corrections. Because of the condition
(\ref{casimir}), we can parametrize the $T_i$'s in terms of two operators,
$z_1$ and $z_2$. As a first step we write:

\begin{equation}
\label{fuzsph}
T_{1}=cosz_1 ~(1-z_2^2)^{\frac{1}{2}},~~~~T_{2}=sinz_1 ~(1-z_2^2)^{\frac{1}{2}}, ~~~~T_{3}=z_2
\end{equation}

These relations should be viewed as defining the operators $z_1$ and
$z_2$. A similar parametrization has been given by T. Holstein and H. Primakoff in terms of creation and annihilation operators \cite{hp}. At the limit of $N$ $\rightarrow$ $\infty$, they become the coordinates
$\phi$ and $cos\theta$ of a unit sphere. To leading order in  $1/N$, the commutation
relations (\ref{rescsu2}) induce the commutation relation (\ref{fuzsphcom}) between
the $z_i$'s: 

\begin{equation}
[z_1,z_2]=\frac{2i}{N}
\end{equation}

\noindent $i.e.$ the $z_i$'s satisfy a Heisenberg commutation relation with the unity
operator at the right-hand side. $1/N$ plays the role of $\hbar$. In higher
orders, however, the definitions (\ref{fuzsph}) must be corrected because the
operators $T_1$ and $T_2$ are no more hermitian. It turns out that  
a convenient choice is to use $T_+$ and $T_-$. We thus write:

\begin{equation}
\begin{split}
\label{fuzsphnew}
T_+ & =T_1+iT_2=e^{\frac{iz_1}{2}}~(1-z_2^2)^{\frac{1}{2}}~e^{
  \frac{iz_1}{2}}\\
T_- & =T_1-iT_2=e^{-\frac{iz_1}{2}}~(1-z_2^2)^{\frac{1}{2}}~e^{-
  \frac{iz_1}{2}}\\
T_3 & =z_2
\end{split}
\end{equation}

The claim is that the $SU(N)$ algebra can be expressed in two equivalent ways:
We can start from the non-commutative coordinates of the fuzzy sphere  $z_1$
and $z_2$ which are assumed to satisfy the quantum mechanical commutation
relations (\ref{fuzsphcom}). Through (\ref{fuzsphnew}) we define
three operators $T_1$, $T_2$ and $T_3$. We shall prove that they satisfy
exactly, without any higher order corrections, the $SU(2)$ relations
(\ref{rescsu2}) and (\ref{casimir}) and, consequently, they can be used as
basis for the entire $SU(N)$ algebra, as it was shown in the previous
section. 
The oposite is also true. The $SU(2)$ commutation relations (\ref{rescsu2}) imply the quantum mechanical
relation (\ref{fuzsphcom}). Starting from the $SU(2)$ generators $T_i$ $i=1,2,3$,
we can use (\ref{fuzsphnew}) to define two operators $z_i$ $i=1,2$. We can
again prove that they satisfy the Heisenberg algebra (\ref{fuzsphcom}) to all orders
in $1/N$. 

We start with the first part of the statement, which is
straightforward calculation. 

We assume (\ref{fuzsphcom}) and we want to compute the commutator of $T_+$ and
$T_-$ given by (\ref{fuzsphnew}).   

\begin{equation}
\begin{split}
\label{proof1}
[T_+,T_-] & =e^{\frac{iz_1}{2}}~(1-z_2^2)~e^{-
  \frac{iz_1}{2}}~-~e^{-\frac{iz_1}{2}}~(1-z_2^2)~e^{\frac{iz_1}{2}}\\
 & =\left(e^{\frac{iz_1}{2}}~z_2~e^{-
  \frac{iz_1}{2}}\right)^2~-~\left(e^{-\frac{iz_1}{2}}~z_2~e^{\frac{iz_1}{2}}\right)^2
\end{split}
\end{equation}

A usefull form of the Cambell-Haussdorf relation for two operators $A$ and $B$
is:

\begin{equation}
\label{CH}
e^A B e^{-A}=B+[A,B]+ \frac{1}{2}[A,[A,B]]+...
\end{equation}

Applying (\ref{CH}) to (\ref{proof1}), we obtain:

\begin{equation}
\label{proof2}
[T_+,T_-]=\left(\frac{1}{N}+z_2 \right)^2~-~\left(\frac{1}{N}-z_2 \right)^2~=~\frac{4}{N}z_2
\end{equation}

Similarly, we check that the other two commutation relations of $SU(2)$ are
satisfied. We can also compute the Casimir operator $T^2$ and we find
the value $1-1/N^2$ of (\ref{casimir}).

We proceed now with the proof of the oposite statement, namely the equivalence
between the $SU(2)$ and the quantum mechanical commutation relations. The essence of the story is that any corrections
on the r.h.s. of (\ref{fuzsphcom}) will affect the $SU(2)$ commutation relations (\ref{rescsu2}). The
argument is inductive, order by order in $1/N$. 

Let us start with the first term and write the general form of (\ref{fuzsphcom}) as:

\begin{equation}
\label{;}
[z_1,z_2]=\frac{1}{N}t_1(z_1,z_2)+O(\frac{1}{N^2})
\end{equation}
with $t_1$ some function of the $z$'s. Using (\ref{;}) we compute the $1/N$ term in the commutator of two $T$'s given by
(\ref{fuzsphnew}). If we assume that they satisfy the $SU(2)$ commutation
relations  we determine
$t_1(z_1,z_2)$:

\begin{equation}
\label{;;;;;;}
t_1(z_1,z_2)=2i
\end{equation}

We can now go back to (\ref{;}) and determine the next term in the
expansion. We write:

\begin{equation}
\label{!}
[z_1,z_2]=\frac{2i}{N}+\frac{1}{N^2}t_2(z_1,z_2)+O(\frac{1}{N^3})
\end{equation}

We look now at the $T_i$'s given by (\ref{fuzsphnew}) and compute  the commutator between $T_+$ and $T_3$ to order
$1/N^2$. Imposing the absence of such terms, we get:

\begin{equation}
\label{!!!!}
t_2(z_1,z_2)=0
\end{equation}

It is now clear how the induction works: We assume the commutator

\begin{equation}
\label{!!!!!}
[z_1,z_2]=\frac{2i}{N}+\frac{1}{N^k}t_k(z_1,z_2)+O(\frac{1}{N^{k+1}})
\end{equation}
and set the coefficient of the corresponding term in the $SU(2)$ commutation
relation equal to zero. This gives again:

\begin{equation}
\label{**}
t_k(z_1,z_2)=0
\end{equation}

\vskip 1cm

The commutation relation (\ref{fuzsphcom}) is the main step of the argument. Any
function $f$ of the $SU(N)$ generators, in particular any polynomial of the Yang-Mills
fields and their space-time derivatives, can be rewritten, using 
 (\ref{fuzsphnew}), as a function of $z_1$ and $z_2$. Since they satisfy the
quantum mechanics commutation relations (\ref{fuzsphcom}), the usual proof of
Moyal \cite{moyal} goes through and the commutator of two such functions $f$ and $g$ will
have an expansion in powers of $1/N$ of the form:

\begin{equation}
\label{moyal}
[f,g] \sim \frac{1}{N} \{f(z_1,z_2),g(z_1,z_2)\} + \frac{1}{N^2}
\left(\{\frac{\partial f}{\partial z_1}, \frac{\partial
  g}{\partial z_2}\} + \{\frac{\partial f}{\partial z_2}, \frac{\partial
  g}{\partial z_1}\} \right) +...
\end{equation}

\noindent with the Poisson brackets defined the usual way: 

\begin{equation}
\label{poisson} 
\{f,g\} = \left(\frac{\partial f}{\partial z_1} \frac{\partial
  g}{\partial z_2} - \frac{\partial f}{\partial z_2} \frac{\partial
  g}{\partial z_1}\right)  
\end{equation}

The first term in this expansion is unambiguous but the coefficients of the
higher orders depend on the particular ordering convention one may adopt. For
example, in the symmetric ordering, only odd powers of $1/N$ appear. 

For the symmetric ordering, we can introduce, formally, a *-product through:

\begin{equation}
\label{star}
f(z)*g(z)=exp(\xi~ \epsilon _{ij}~\partial _z^i \partial _w^j)f(z)g(w)|_{w=z}
\end{equation}

\noindent with $z=(z_1,z_2)$ and $\xi=\frac{2i}{N}$. The $SU(N)$ commutators
in the Yang-Mills Lagrangian can now be replaced by the *-products on the
non-commutative surface. This equality will be exact at any given order in the
$1/N$ expansion. This completes the proof of our statement. Let us remark here that the change of variables going from $SU(N)$  gauge potentials to functions
over the noncommutative sphere for every N, is an invertible mapping
and it is the one which induces from the  matrix product
the star (Moyal) product. At the classical level this induces a correspondance between the two theories. If we want to consider the quantum theory, we must also compute the Jacobian of the tranformation for the measure of the functional integration. At the limit $N \rightarrow \infty$ we obtain an ordinary $d+2$-dimensional field theory with the two dimensions forming a commuting $S_2$.

\section{Conclusions}

Although our proof has been given for the case of the sphere, it can
be extended to  surfaces with higher genus \cite{novikov}, \cite{jaffe}. 

It is straightforward to generalize these results and include matter fields,
provided they also belong to the adjoint representation of $SU(N)$. In
particular, the supersymmetric Yang-Mills theories have the same property. The
special case of ${\cal N}=4$ supersymmetry is of obvious interest because of
its conformal properties. In this theory the duality $g \rightarrow 1/g$ makes
us believe that 
 the two large $N$ limits, namely (\ref{hooftlim}) and (\ref{newlimit}), are related.

 We believe that one could also
include fields belonging to the fundamental representation of the gauge
group. In 't Hooft's limit such matter multiplets are restricted to the
edges of the diagram, so we expect in our case the generalization to
involve open surfaces.

The equivalence between the original $d$-dimensional Yang-Mills theory and the
new one in $d+2$ dimensions holds at the classical level. For the new
formulation however, the ordinary perturbation series, even at the large $N$ limit, is divergent. The reason is that the quadratic
part of this action does not contain derivatives with respect to
$z_1$ or $z_2$. This is not surprising because these
divergences represent the factors of $N$ in the diagrams of the
original theory which, contrary to what happens in (\ref {hooftlim}), have
not been absorbed in the redefinition of the coupling constant.
However, we expect a perturbation expansion around some appropriate
non-trivial classical solution to be meaningful and to contain
interesting information concerning the strong coupling limit of the
original theory. 

A final remark: Could one have anticipated the emergence of this action in the $1/N$ expansion? Coming back to (\ref {lint}), we can
always replace $\phi^i(x)$ $i=1,...,N$, at the limit when $N$ goes to
infinity, with $\phi (\sigma, x)$ where $0 \leq \sigma \leq 2 \pi
$. In this case the sum over i will become an integral over
$\sigma $. However, the $\phi^4$ term in (\ref {lint}) will no more be local in
$\sigma $. So, the only surprising feature  is that,
for a Yang-Mills theory, the resulting expression is local.

\vskip 3cm

{\large \bf Acknowledgments}

One of us, (J. I.), wishes to thank Prof. C. Bachas and E. Kiritsis for clarifying discussions.

\end{document}